\documentclass[fleqn,usenatbib,twocolumn]{aastex631}
\usepackage{newtxtext}
\usepackage{newtxmath}
\usepackage[T1]{fontenc}
\usepackage{graphicx}
\usepackage{amsmath}
\usepackage{amssymb}
\usepackage{upgreek}
\usepackage{enumerate}
\usepackage{threeparttable}

\shorttitle{Serendipitous discovery of a new radio pulsar}
\shortauthors{Xie et al. }

\usepackage{soul}

\newcommand{\del}[1]{\textcolor{magenta}{{\iffalse{#1}\fi}}}

\begin{document}

\title{Searching radio signals from two magnetars and a high-magnetic field pulsar and the serendipitous discovery of a new radio pulsar PSR J1935+2200}

\correspondingauthor{J.L. Han}
\email{hjl@nao.cas.cn}

\author[0000-0003-1946-086X]{Lang Xie}
\affiliation{National Astronomical Observatories, Chinese Academy of Sciences, Beĳing 100101, China}
\affiliation{School of Astronomy, University of Chinese Academy of Sciences, Beĳing 100049, China}

\author[0000-0002-9274-3092]{J. L. Han}
\affiliation{National Astronomical Observatories, Chinese Academy of Sciences, Beĳing 100101, China}
\affiliation{School of Astronomy, University of Chinese Academy of Sciences, Beĳing 100049, China}

\author[0009-0009-6590-1540]{Z.L. Yang}
\affiliation{National Astronomical Observatories, Chinese Academy of Sciences, Beĳing 100101, China}
\affiliation{School of Astronomy, University of Chinese Academy of Sciences, Beĳing 100049, China}

\author[0000-0002-1056-5895]{W.~C. Jing}
\affiliation{National Astronomical Observatories, Chinese Academy of Sciences, Beĳing 100101, China}
\affiliation{School of Astronomy, University of Chinese Academy of Sciences, Beĳing 100049, China}

\author[0000-0002-6423-6106]{D.~J. Zhou}
\affiliation{National Astronomical Observatories, Chinese Academy of Sciences, Beĳing 100101, China}
\affiliation{School of Astronomy, University of Chinese Academy of Sciences, Beĳing 100049, China}

\author[0009-0003-2212-4792]{W.~Q. Su}
\affiliation{National Astronomical Observatories, Chinese Academy of Sciences, Beĳing 100101, China}
\affiliation{School of Astronomy, University of Chinese Academy of Sciences, Beĳing 100049, China}

\author[0009-0008-1612-9948]{Yi Yan}
\affiliation{National Astronomical Observatories, Chinese Academy of Sciences, Beĳing 100101, China}
\affiliation{School of Astronomy, University of Chinese Academy of Sciences, Beĳing 100049, China}

\author[0000-0002-4704-5340]{Tao Wang}
\affiliation{National Astronomical Observatories, Chinese Academy of Sciences, Beĳing 100101, China}
\affiliation{School of Astronomy, University of Chinese Academy of Sciences, Beĳing 100049, China}

\author[0000-0002-5915-5539]{Nan-Nan Cai}
\affiliation{National Astronomical Observatories, Chinese Academy of Sciences, Beĳing 100101, China}
\affiliation{School of Astronomy, University of Chinese Academy of Sciences, Beĳing 100049, China}

\author[0000-0002-6437-0487]{P.~F. Wang}
\affiliation{National Astronomical Observatories, Chinese Academy of Sciences, Beĳing 100101, China}
\affiliation{School of Astronomy, University of Chinese Academy of Sciences, Beĳing 100049, China}

\author[0009-0004-3433-2027]{Chen Wang}
\affiliation{National Astronomical Observatories, Chinese Academy of Sciences, Beĳing 100101, China}
\affiliation{School of Astronomy, University of Chinese Academy of Sciences, Beĳing 100049, China}

\begin{abstract}
Magnetars are slowly rotating, highly magnetized young neutron stars that can show transient radio phenomena for radio pulses and fast radio bursts. We conducted radio observations of from two magnetars  SGR$~$J1935+2154 and 3XMM$~$J185246.6+003317 and a high-magnetic field pulsar PSR$~$J1846$-$0258 using the Five-hundred-meter Aperture Spherical radio Telescope (FAST). We performed single pulse and periodicity searches and did not detect radio signals from them. From the piggyback data recorded by other FAST telescope beams when we observed the magnetar SGR$~$1935+2154, we serendipitously discovered a new radio pulsar, PSR$~$J1935+2200. We carried out the follow-up observations and obtained the timing solution based on these new observations and the archive FAST data. PSR$~$J1935+2200 is an isolated old pulsar, with a spin period of $0.91$s, a spin-period derivative of $9.19 \times 10^{-15}$~s~s$^{-1}$, and a characteristic age of $1.57$ Myr. It is a weak pulsar with a flux density of 9.8 $\mu$Jy at 1.25 GHz. Discovery of a new pulsar from the long FAST observations of 30 minutes implies that there may be more weak older pulsars in the Galactic disk to be discovered.
\end{abstract}


\section{Introduction} \label{sec:intro}

Magnetars are slowly rotating young neutron stars with super-strong magnetic fields ($B>10^{14}$\mbox{ G}) \citep{Mereghetti2015,Kaspi2017}, which are known for their widely diverse X-ray activities, such as short bursts, sustained outbursts, burst storms and Giant Flares \citep{Rea2010}. The energy associated with such activities can be interpreted as the decay of the magnetic field \citep{Duncan1992}.
Up to now, only six of nearly 30 magnetars have radio emission detected\footnote{http://www.physics.mcgill.ca/~pulsar/magnetar/main.html}, namely, 1E$~$1547.0-5408, Swift$~$J1818.0-1607, SGR$~$1745-2900, PSR$~$J1622-4950, XTE$~$J1810-197 and SGR$~$1935+2154 \citep{Olausen2014}. Radio emission from magnetars is often temporally associated with X-ray outbursts \citep{Camilo2006, Camilo2007a}. Radio pulses from magnetars have a strong variability in flux density and also profile shape compared to rotating powered pulsars \citep{Kramer2007,Camilo2007a,Levin2010,Kirsten2021}, which is possibly caused by untwisting of the dynamic magnetosphere. 
In addition, the averaged pulse profiles and single pulses often have highly linear polarisation \citep{Lower2020}. The radio spectrum tends to be flat or inverted \citep{Camilo2007a}.

\begin{table*}
\caption{Flux density limits of the periodicity search and single pulse search of three sources}
\begin{center}
\begin{tabular}{lcccccccccl}
\hline
\hline
Object & P  & $\dot{P}$ & B & DM  & FAST Obs.Date & T$_{\rm obs}$ &$S_{\rm periodic}$    &$S_{\rm single}$&   \\
&  (s)  & ($10^{-11}$~s~s$^{-1}$) & ($10^{14}$G)  & (pc cm$^{-3}$) & (yyyymmdd) & (s)  & ($\mu$Jy) &  (mJy)  \\
\hline
SGR J1935+2154  & 3.24 & 1.43 & 2.2 &293.69  & 20240408 &1800  & 1.7  & 16.5 \\
PSR J1846$-$0258 & 0.324 & 0.7 & 0.49 & 317.51 & 20240221  &1200  & 2.1  & 17.0      \\
                &  &  &              &    & 20240408  &1200  &   &     \\
3XMM J185246.6+003317 & 11.56 &   $<0.014$ &  	$<0.41$ &441.46  & 20240408 &600   & 2.9 & 18.9     
\\
\hline
\end{tabular}
 \begin{tablenotes}   
        \footnotesize               
        \item[1] Note: DM values are estimated by using the Galactic electron density model NE2001 \citep{Cordes2002}. $S_{\rm periodic}$ and $S_{\rm singular}$ are the 7$\sigma$ for periodicity and single pulse search, respectively.                  
      \end{tablenotes}     
\end{center}
\label{table:limits}
\end{table*}

SGR$~$1935+2154 was discovered in 2014 July by Swift/BAT via its X-ray short bursts \citep{Stamatikos2014}. Its position is associated with SNR$~$G57.2+0.8 \citep{Gaensler2014}, which corresponds to a distance in the range of $6.6$-$12.5$ kpc \citep{Sun2011, Zhong2020,Zhou2020,Kirsten2021}.
Subsequent observations by Chandra and XMM-Newton Telescopes have revealed the continuous X-ray counterpart of SGR 1935+2154 with a period of $P = 3.24$\mbox{ s}, a spin-down rate of $\dot{P} = 1.43\times10^{-11}$~s~s$^{-1}$, which corresponds to a characteristic age of 3.6 kyr and a dipolar magnetic field of $B =  2.2 \times 10^{14}$\mbox{ G} \citep{Israel2016}. 
On 2020 April 28, the Canadian Hydrogen Intensity Mapping Experiment (CHIME) detected the millisecond duration radio burst (FRB) 20200428 associated with an X-ray burst from the SGR$~$1935+2154 \citep{CHIME2020,Li2021}, which proved the link between FRBs and magnetars.
Follow-up observations reveal the activity of radio emission from SGR$~$1935+2154 \citep{Kirsten2021, Hu2024}. The Five-hundred-meter Aperture Spherical radio Telescope  \citep[FAST,][]{Nan2006,Nan2011} also detected the radio pulse radiating from SGR$~$1935+2154 \citep{Zhu2023}.
Many open questions remain about the connection between X-ray bursts, pulsar-like radio emissions and FRBs and their physical origin.

PSR$~$J1846$-$0258 is a young, highly magnetic pulsar exhibiting magnetar-like X-ray bursts, making it an intermediate object between magnetars and radio pulsars \citep{Gavriil2008}. PSR$~$J1846$-$0258 is probably a high magnetic field pulsar located at the center of the SNR Kes75 at a distance of 5.8 kpc \citep{Leahy2008}, and has been historically radio-quiet. X-ray observations show a pulsation period of $P = 0.324$\mbox{ s}, a dipole magnetic field of $B = 4.9 \times 10^{13}$\mbox{ G}, and a characteristic age of 723 years. Between May and June of 2020, Swift/BAT and NICER detected the source experiencing another magnetic-like X-ray burst \citep{Krimm2020,Sathyaprakash2024}. No radio pulses were detected by follow-up observations \citep{Majid2020}, though it is probably a rotation-powered pulsar most of the time.

3XMM$~$J185246.6$+$003317 (hereafter 3XMM J1852+0033) is a low magnetic field magnetar discovered by \citet{zhou2014} by using the XMM-Newton data, and it is located near SNR Kes79 and a distance of 7.1 kpc \citep{Rea2014}. A search of the XMM-Newton archive shows that it experienced intense X-ray activity between 2008 and 2009. 
3XMM$~$J1852+0033 has a spin period of $P = 11.56$\mbox{ s} and a spin-down rate $\dot{P} < 1.4\times10^{-13}$~s~s$^{-1}$, which implies a characteristic age of $\tau> 1.3 $\mbox{ Myr} and a dipolar magnetic field of $B < 4.1 \times 10^{13}$\mbox{ G} \citep{Rea2014}. No further outbursts have been reported and no radio signals have been detected so far.
 
The FAST \citep{Nan2006,Nan2011} is currently the most sensitive telescope in the world \citep{Jiang2020} and is also a powerful tool for searching for radio signals from magnetars and faint pulsars. Although the sky areas where these magnetars are located have been searched for pulsars many times by many radio telescopes, the super-sensitive FAST could still find weak pulsars given the longer time of tracking observations. The weak pulsars are fundamental for the lower end of the pulsar luminosity function, and potentially important for our understanding of extreme objects in the universe and stellar evolution if they are in binary system.

Previously, radio signals from SGR$~$J1935+2154 have been successfully detected, and prompted many follow-up observations \citep{Good2020,Lin2020, Zhang2020,Bailes2021, Kirsten2021, Tang2021, Lu2024}.
We are motivated to conduct radio observations of these from two magnetars and a high-magnetic field pulsar 
(SGR$~$J1935+2154, 3XMM$~$J1852+0033 and PSR$~$J1846$-$0258) by using FAST mainly for possible radio signals or get a strict upper limit on the flux density. We describe the observations and data processing in Section \ref{sec2}. Section \ref{sec3} presents the results and we discuss and summarize in Section \ref{sec4}.

\section{FAST Observations And Data Reduction} \label{sec2}

We carried out observations of SGR$~$J1935+2154, PSR$~$J1846-0258 and 3XMM$~$J1852+0033 using FAST (project PT2023\_0176, PI: Lang Xie) in the four sessions: 
(\romannumeral1) 2024 February 21 9:16:00 to 9:36:00 UTC for PSR$~$J1846$-$0258; (\romannumeral2) 2024 April 8 6:20:00 to 6:40:00 UTC for PSR$~$J1846$-$0258; (\romannumeral3) 2024 April 8 6:52:00 to 7:02:00 UTC for 3XMM$~$J1852+0033; 
(\romannumeral4); 2024 April 8 7:14:00 to 7:44:00 UTC for SGR$~$ J1935+2154.
All observations were performed in the tracking mode with data from the $L$-band 19-beam receiver recorded in the pulsar search format. The receiver covers the frequency range of $1.0$-$1.5$ GHz with 2048 frequency channels. For each channel, the data of four polarization channels of  $XX$, $YY$, Re[$X^{*}Y$] and Im[$X^{*}Y$] have been recorded in fits files with a sampling time of 49.152$\mu$s. At the beginning of each observation session, the calibration signals with a 1~K equivalent white noise and a 2 s period are injected into the feed, and the data are recorded for system calibration. The three sources are tracked by using the FAST $L$-band central beam. During observations, the data from the other 18 beams are also recorded. 

We search for radio pulses using the PRESTO module \citep{Ransom2011} and the single-pulse search module \citep{Zhou2023}, both of which have been used for the Galactic Plane Pulsar Snapshot (GPPS) survey \citep{Han2021,Han2025}. Data processing includes the following steps: (1) Use \textit{rfifind} to detect radio frequency interference (RFI) and then create a mask file to exclude data contaminated by RFI. (2) De-dispersed data from all frequency channels using \textit{prepsubband} based on the estimated DM values. (3) Search for periodic signals is performed using \textit{accelsearch}. (4) Confirm candidates and fold data using \textit{prepfold}.

Based on the timing results of X-ray observations, we can obtain the initial positions and periods of the three  sources.  We use the \textit{rfifind} to search for RFI and create the mask. We perform the standard dedispersion procedure as done in the FAST GPPS survey \citep{Han2021}, with fine DM steps of 0.1, 0.15, 0.3, 0.5 pc\,cm$^{-3}$ in the DM range of 5 -- 1350 pc\,cm$^{-3}$. Subsequently, we performed a periodic search of the observations of these two magnetars and a high-magnetic field pulsar using the derived spin periods.

In addition, the single pulse search module is used to search for radio pulses \citep{Zhou2023}. First, we de-disperse the data between 3-1000 pc \,cm$^{-3}$ with a step size of 1 pc \,cm$^{-3}$, and then generate DM-time images with every 4 s of data. Then we search for salient points in the images of DM-time using the YOLO target detection technology developed in the Darknet neural network framework \citep{Bochkovskiy2020a}. Finally, we use the artificial intelligence (AI) technology to quickly identify individual pulses, and perform period finding for the collected pulses.

\section{RESULTS} \label{sec3}

Pulse searches have null results for two magnetars and one high-magnetic-field pulsar by using both the period folding and the single pulse search module,
with a signal-to-noise ratio threshold of S/N $>$ 7.
But a new radio pulsar was discovered serendipitously in the M07 beam when we observed SGR$~$J1935+2154.

\begin{table}[]
    \centering
    \caption{FAST observations of PSR J1935+2200. Data obtained before 20240408 are released archive data.}
    \small
    \begin{tabular}{clhhcc}
    \hline
 Obs. Date  & FAST Beam Name & Project ID & PI & T$_{\rm obs}$  & Beam Offset \\
 (yyyymmdd) & & & & (s) & (arcmin)\\
 \hline
 20191106 & SGR1935+2154: M07 & 2019a-132-O & Bing Zhang & 10800 & 1.5\\
 20191107 & SGR1935+2154: M07 & 2019a-132-O & Bing Zhang & 10800 & 1.5 \\
 20200430 & SGRJ1935+2154: M07 & ZD2020\_5	& Bing Zhang & 3840 & 1.5\\
 20200501 & SGRJ1935+2154: M07 & ZD2020\_5	& Bing Zhang & 3840 & 1.5\\
 20200507 & SGRJ1935+2154: M07 & ZD2020\_5	& Bing Zhang & 3600 & 1.5\\
 20200508 & SGRJ1935+2154: M07 & ZD2020\_5	& Bing Zhang & 5400 & 1.5\\
 20200509 & SGRJ1935+2154: M07 & ZD2020\_5	& Bing Zhang & 5400& 1.5\\
 20200513 & SGRJ1935+2154: M07 & ZD2020\_5	& Bing Zhang & 4200& 1.5\\
 20200515 & SGRJ1935+2154: M07 & ZD2020\_5	& Bing Zhang & 5400& 1.5\\
 20200517 & SGRJ1935+2154: M07 & ZD2020\_5	& Bing Zhang & 5400& 1.5\\
 20201204 & G57.25+0.93: P2M09 & ZD2020\_2	& J.~L. Han & 300& 1.5\\
 20240408 & SGR1935+2154: M07 & PT2023\_0176 & Xie lang & 1800& 1.5\\
 20240714 & J193508+215846: M01 & ZD2023\_2 & J.~L. Han & 600 & 1.5\\
 20240917 & J193506+J220016: M01 & ZD2024\_2 & J.~L. Han & 900 & 0.0\\
 \hline
    \end{tabular}
    \label{tab:obs}
\end{table}

\subsection{Upper Limits of flux densities }

No radio bursts from three sources are detected by FAST, which suggests the upper limits on the flux densities of three sources for radio radiation in their normal state.
We can derive an upper limit on the flux density of the pulsar periodicity search based on the modified radiometric equations \citep{Lorimer2004}
\begin{equation}\label{1}
    S_{\rm periodic} = \frac{ 
    (S/ N)_{\rm min} T_{\rm  sys}}{G_0 \cdot \sqrt {n_p \cdot BW \cdot t_{\rm obs} }}\sqrt{\frac{W_{\rm obs}}{P - W_{\rm obs}}} \ \ \ 
\end{equation}
where the system noise temperature is $T_{\rm  sys} = 22~$K  
\citep{Jiang2020}, the effective gain of the telescope is $G_0 = 16.1$ K/Jy, the frequency bandwidth is $BW$ = 437.5 MHz (after interference is removed), the number of polarisation channels is $n_p=2$, 
the integration time is $t_{\rm obs}$ and 
the observed pulse width $W_{\rm obs}$ can be estimated by 
$W_{\rm obs}  = \sqrt{ W_{\rm intrinsic}^2 + t_{\rm bin}^2 + \tau_{s}^2 + \Delta t_{\rm chan}^2}$, here $W_{\rm intrinsic}$ is intrinsic pulse width, $t_{\rm bin}$ is the sampling time, 
$\tau_{s}$ is the scattering timescale and $\Delta t_{\rm chan}$ is the dispersion time inside one frequency channel.
Assuming a pulsar duty cycle of 0.03 and taking the minimum detection threshold S/N $=$ 7, our non-detection of signals sets 
flux density limits of periodicity searches 
for two magnetars and a high-magnetic field pulsar given in Table \ref{table:limits}.

For a single pulse search, the sensitivity limit of the flux density can be described as:
\begin{equation}\label{fluxsingle}
    S_{\rm single} =\frac{W_{\rm obs}} {W_{\rm intrinsic}} \frac{ 
    (S/N)_{\rm min}  T_{\rm sys}  }   {G_0 \cdot \sqrt {n_p \cdot BW \cdot W_{\rm obs} } }.
\end{equation}
Assuming that the intrinsic width of the pulse is 1~ms, one can get the upper limits of 7$\sigma$ for the flux densities of from two magnetars and a high-magnetic field pulsar as listed in Table \ref{table:limits}. 

\begin{figure}
 \centering
\includegraphics[width=0.42\textwidth]{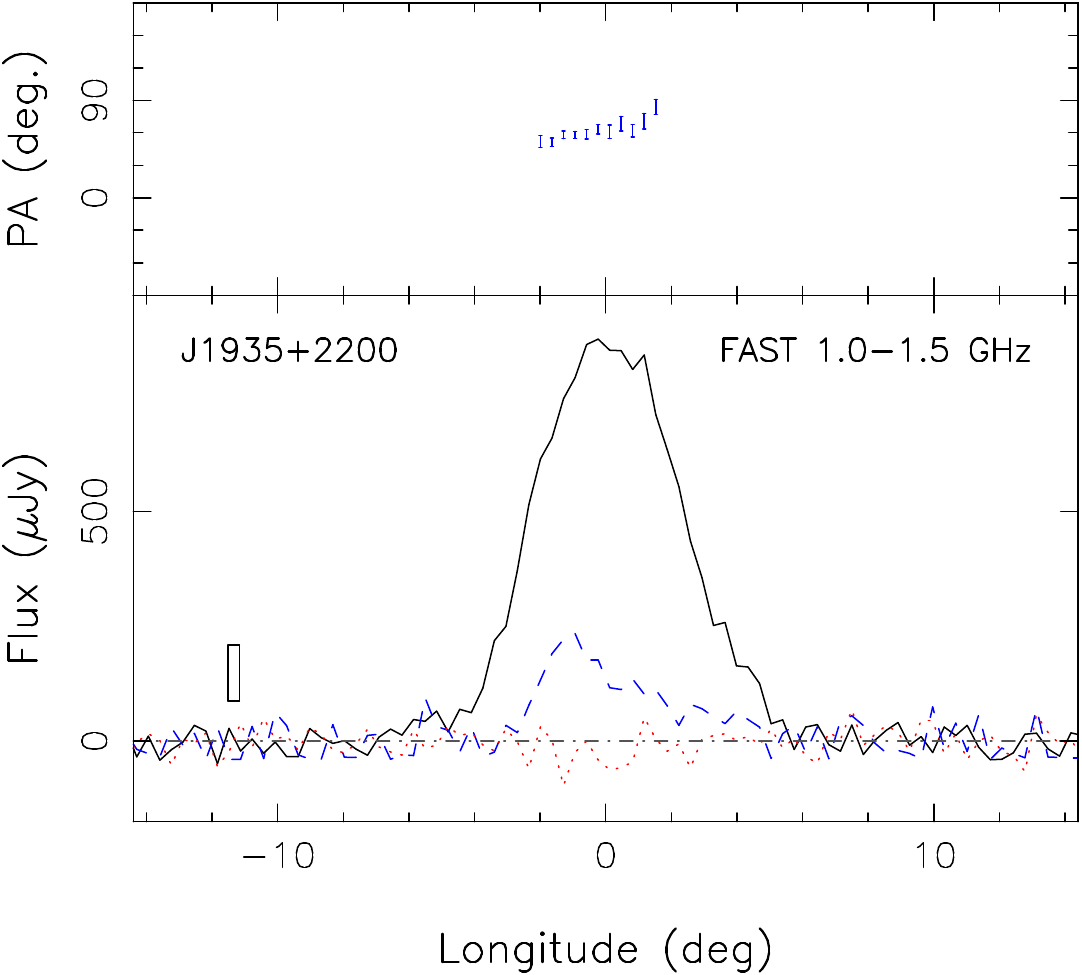}
\caption {Polarization profile of PSR J1935+2200. The total intensity (solid line), linear polarization (dashed line), and circular polarization (dashed line) are shown in the {\it lower subpanel} with a bix indication for $\pm1\sigma$ and bin size, and polarization angles are shown in the upper subpanel. }
\label{Polarization profile}
\end{figure}

\begin{table}
\centering
\caption{Observation parameters and measured and derived parameters for PSR J1935+2200.}
\label{timingsolution}
\small
\begin{tabular}{lc}
\hline
Parameters & Values \\
\hline
Timing span (MJD) & 58793-60570\\
Solar system ephemeris model & DE436\\
Time units & TCB\\
Number of TOAs & 28\\
Epoch of timing solution (MJD) & 59808.0000\\
Clock correction procedure & TT(TAI)\\
R.A., $\alpha_\mathrm{J2000}$ & $19^{\mathrm{h}}35^{\mathrm{m}}06\fs983(5)$\\
decl., $\delta_\mathrm{J2000}$ & $22\degr00\arcmin15\farcs947(6)$\\
Galactic longitude (deg) & 57.70435\\ 
Galactic latitude (deg) & 0.45317\\
Spin period (s) & 0.912005933(1)\\
Spin period derivative ($10^{-15} {\rm s\;s}^{-1}$) & 9.19483(1)\\
Weighted post-fit residual ($\upmu$s) & 392.403\\
Characteristic age (Myr) & 1.57\\
Surface dipole magnetic field strength (G) & $2.93 \times 10^{12
}$\\
Spin-down luminosity (erg s$^{-1}$) & $1.83 \times 10^{32}$\\
Dispersion Measure, $\mathrm{DM}$ (pc cm$^{-3}$) & 293.68(2)\\
$\mathrm{DM}$ distance (kpc for NE2000 / YMW16) & 8.8 / 9.0\\
Rotation Measure, $\mathrm{RM}$ (rad m$^{-2}$) & 254(3)\\
Flux density ($\mu$Jy at 1250~MHz) & 9.8(6) \\
Spectral index of flux density   & $-2.1(6)$ \\
\hline
\end{tabular}
\end{table}

\begin{figure} 
 \centering
\includegraphics[width=0.42\textwidth]{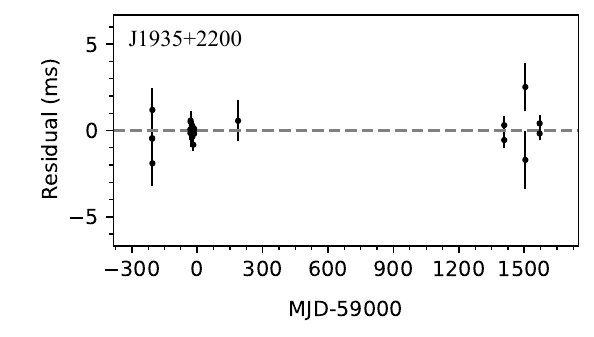}
\caption {The post-fit timing residuals of PSR J1935+2200 for the phase-connected timing solution. }
\label{timing}
\end{figure}

\subsection{Serendipitous Discovery}

A new radio pulsar PSR J1935+2200 was serendipitously discovered when we observed the magnetar SGR 1935+2154 on 2024 April 8 using the FAST. The pulsar stands out in the periodical pulse search 
in M07 beam. Re-processing of the released archived FAST data on the same position confirmed the detection of the pulsar (see Table~\ref{tab:obs}). 

After the initial detection of the pulsar, we conducted 2 follow-up observations. The data were de-dispersed by using the best DM values obtained earlier and then folded. 
PSR J1935+2200 can be found by using the DSPSR software package \citep{van2011}. 
We use the {\it PSRCHIVE} to remove the frequency channels affected by RFI \citep{Hotan2004}, integrate data from all frequency channels, and then use {\it PAAS} to create a noise-free standard profile template to determine the time of arrival (TOA).

\begin{figure}
  \centering
  \includegraphics[width=0.45\textwidth]{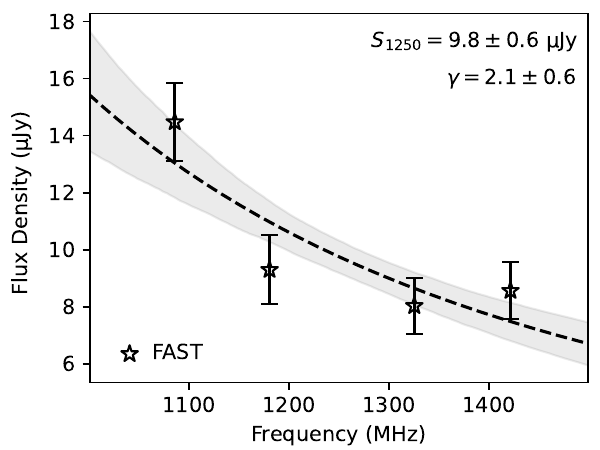}
  \caption{Flux densities of PSR J1935+2200 in the 4 subbands in the frequency range of 1.0 -- 1.5 GHz observed by FAST and the best fitting for the power-law index $\gamma$ for $S_{\nu} \sim \nu^{-\gamma}$. The $\pm1\sigma$ uncertainties are given for flux densities and the confidence level of the fitting in the shadow area.}
\label{flux}
\end{figure}

By using all FAST data listed in Table~\ref{tab:obs}, newly observed and previously archived, 
we obtained the phase-coherent timing solution of this pulsar, as given in Table~\ref{timingsolution}, and the polarization profiles given in Figure \ref{Polarization profile}.  The timing residuals are shown in Figure \ref{timing}. PSR J1935+2200 has a period of 0.91 s and a period derivative of $9.19 \times 10^{-15}$ s s$^{-1}$, which corresponds to a characteristic age of 1.57 \mbox{ Myr}, a surface magnetic field strength of $2.93 \times 10^{12 }$\mbox{ G}, and a spin decay luminosity of $1.83 \times 10^{32}$ erg s$^{-1}$. It is a slightly old pulsar as indicated by the characteristic age. Based on currently limited timing data, we conclude that it is an isolated pulsar, otherwise must be in a binary system with an extremely long orbit period or an extremely light companion, which we could not measure from currently available timing data.

With the high time resolution of FAST, we can determine the optimal DM value to be 293.68(2) pc cm$^{-3}$. 
The distance of this pulsar is estimated to be 8.8 or 9.0 kpc according to the Galactic electron density distribution models, NE2001 and YMW16 \citep{Cordes2002, Yao2017}. 

Because we have the 4-polarization channel data, we obtained the polarization profile of PSR J1935+2200, as shown in Figure~\ref{Polarization profile}. 
Polarization was calibrated using 
using PAM in the PSRCHIVE software \citep{Hotan2004}.
We measured pulse-averaged flux density of 9.8(6) $\mu$Jy at a center frequency of 1250 MHz, as shown in Figure \ref{flux} and also the spectral index $\gamma = 2.1(6)$ from the flux densities of four subbands.

\section{DISCUSSION AND  CONCLUSION} \label{sec4}

We searched for radio periodic signals and single pulses from two magnetars and one high-magnetic-field pulsar with FAST, and did not detect radio pulses with dispersive features and signal-to-noise ratios of S/N $>$ 7. The observations provide the upper limit on the flux density for single pulse and periodicity searches of the three sources. 
Previously, \cite{Zhang2020} captured a radio event from SGR$~$1935+2154 on 2020 April 30 using FAST with a fluence of 51(2) mJy~ms. 
\citet{Kirsten2021} detected two moderately bright radio bursts by using the Westerbork telescope on 2020 May 24 with fluxes of 112(22) Jy ms and 24(5) Jy ms (22), respectively. \citet{Lin2020} used FAST to observe SGR$~$1935+2154 for 8 hours and gave an upper limit of the radio flux as 22 mJy ms, which is close to our result.

Radio bursts from magnetars are relatively rare transient phenomena. 
Two main hypotheses have been proposed regarding the mechanism of radio emission from magnetars. One is similar to the traditional model of pulsar radio emission and suggests that the radio emission arises from the outflow of highly relativistic particles from the polar cap region of the magnetosphere \citep{Ruderman1975, Arons1979, Cheng1986}. The other holds that the radio emission originates from the J-bundle in the closed magnetosphere, which maintains fluctuations in the net charge through a process of untwisting of the magnetic lines, thus providing conditions for particle acceleration and radio-coherent radiation \citep{Duncan1992, Beloborodov2009, Wang2019}. 
Radio pulses from from two magnetars and a high-magnetic field pulsar were not detected by FAST, which provides important constraints on the physical mechanisms by which magnetars produce radio bursts. Based on the two types of radio emission models described above, we suggest that a variety of factors may work together to quench 
radio emission from most magnetars.
The first possibility is that the magnetar may be in an X-ray quiescence and cannot fulfill the conditions for the formation of J-bundle. 
The second possibility is that the magnetar is unable to produce a large number of electron-positron pairs in some special cases. Some studies have shown that if the strength of the magnetic field exceeds the quantum critical field, the $\gamma\gamma$ QED processes would be strongly suppressed \citep{Daugherty1996,Hibschman2001}, which largely leads to the ``radio quiet''.
The last possibility is that the magnetar's radio beams are more collimated than the high-energy jets, and most of them do not pass through the Earth. 

We report the discovery of a new radio pulsar, PSR J1935+2200, and obtain timing solutions based on previously available and newly conducted 
observations by FAST. 
We find that PSR J1935+2200 is an isolated old pulsar with a spin period of 0.91\mbox{ s} and a spin period derivative of $9.19 \times 10^{-15}$~s~s$^{-1}$. 
Using the high temporal resolution of FAST, we can determine the optimal DM value of 293.68(2) pc cm$^{-3}$. The measured DM values indicate distances of 8.8 and 9.0 kpc for PSR J1935+2200 according to the NE2001 and YMW16 models, respectively.  We searched the archived images from the Swift, Chandra, and XMM-Newton telescopes and found no X-ray or optical counterpart of this object. This pulsar was shown up by FAST integration of 30 minutes but was missed in the survey due to the beam offset, suggesting that there may be more old pulsars with weak radio emission in the Milky Way disk to discover. 

\begin{acknowledgments}
We thank the anonymous referee for helpful comments. This work made use of the data from FAST (http://cstr.cn/31116. 02.FAST). FAST is a Chinese national mega-science facility, built and operated by the National Astronomical Observatories, Chinese Academy of Sciences. We all appreciate the excellent performance of the FAST and the operation team. The authors are supported by the National Natural Science Foundation of China: Nos. 11988101, and 11833009, and also the National SKA Program of China 2020SKA0120100.
\end{acknowledgments}

\bibliography{ms}

\begin{thebibliography}{}
\expandafter\ifx\csname natexlab\endcsname\relax\def\natexlab#1{#1}\fi
\providecommand{\url}[1]{\href{#1}{#1}}
\providecommand{\dodoi}[1]{doi:~\href{http://doi.org/#1}{\nolinkurl{#1}}}
\providecommand{\doeprint}[1]{\href{http://ascl.net/#1}{\nolinkurl{http://ascl.net/#1}}}
\providecommand{\doarXiv}[1]{\href{https://arxiv.org/abs/#1}{\nolinkurl{https://arxiv.org/abs/#1}}}

\bibitem[{{Arons} \& {Scharlemann}(1979)}]{Arons1979}
{Arons}, J., \& {Scharlemann}, E.~T. 1979, \apj, 231, 854, \dodoi{10.1086/157250}

\bibitem[{{Bailes} {et~al.}(2021){Bailes}, {Bassa}, {Bernardi}, {Buchner}, {Burgay}, {Caleb}, {Cooper}, {Desvignes}, {Groot}, {Heywood}, {Jankowski}, {Karuppusamy}, {Kramer}, {Malenta}, {Naldi}, {Pilia}, {Pupillo}, {Rajwade}, {Spitler}, {Surnis}, {Stappers}, {Addis}, {Bloemen}, {Bezuidenhout}, {Bianchi}, {Champion}, {Chen}, {Driessen}, {Geyer}, {Gourdji}, {Hessels}, {Kondratiev}, {Klein-Wolt}, {K{\"o}rding}, {Le Poole}, {Liu}, {Lower}, {Lyne}, {Magro}, {McBride}, {Mickaliger}, {Morello}, {Parthasarathy}, {Paterson}, {Perera}, {Pieterse}, {Pleunis}, {Possenti}, {Rowlinson}, {Serylak}, {Setti}, {Tavani}, {Wijers}, {ter Veen}, {Venkatraman Krishnan}, {Vreeswijk}, \& {Woudt}}]{Bailes2021}
{Bailes}, M., {Bassa}, C.~G., {Bernardi}, G., {et~al.} 2021, \mnras, 503, 5367, \dodoi{10.1093/mnras/stab749}

\bibitem[{{Beloborodov}(2009)}]{Beloborodov2009}
{Beloborodov}, A.~M. 2009, \apj, 703, 1044, \dodoi{10.1088/0004-637X/703/1/1044}

\bibitem[{{Bochkovskiy} {et~al.}(2020){Bochkovskiy}, {Wang}, \& {Liao}}]{Bochkovskiy2020a}
{Bochkovskiy}, A., {Wang}, C.-Y., \& {Liao}, H.-Y.~M. 2020, arXiv e-prints, arXiv:2004.10934, \dodoi{10.48550/arXiv.2004.10934}

\bibitem[{{Camilo} {et~al.}(2006){Camilo}, {Ransom}, {Halpern}, {Reynolds}, {Helfand}, {Zimmerman}, \& {Sarkissian}}]{Camilo2006}
{Camilo}, F., {Ransom}, S.~M., {Halpern}, J.~P., {et~al.} 2006, \nat, 442, 892, \dodoi{10.1038/nature04986}

\bibitem[{{Camilo} {et~al.}(2007){Camilo}, {Cognard}, {Ransom}, {Halpern}, {Reynolds}, {Zimmerman}, {Gotthelf}, {Helfand}, {Demorest}, {Theureau}, \& {Backer}}]{Camilo2007a}
{Camilo}, F., {Cognard}, I., {Ransom}, S.~M., {et~al.} 2007, \apj, 663, 497, \dodoi{10.1086/518226}

\bibitem[{{Cheng} {et~al.}(1986){Cheng}, {Ho}, \& {Ruderman}}]{Cheng1986}
{Cheng}, K.~S., {Ho}, C., \& {Ruderman}, M. 1986, \apj, 300, 500, \dodoi{10.1086/163829}

\bibitem[{{CHIME/FRB Collaboration} {et~al.}(2020){CHIME/FRB Collaboration}, {Andersen}, {Bandura}, {Bhardwaj}, {Bij}, {Boyce}, {Boyle}, {Brar}, {Cassanelli}, {Chawla}, {Chen}, {Cliche}, {Cook}, {Cubranic}, {Curtin}, {Denman}, {Dobbs}, {Dong}, {Fandino}, {Fonseca}, {Gaensler}, {Giri}, {Good}, {Halpern}, {Hill}, {Hinshaw}, {H{\"o}fer}, {Josephy}, {Kania}, {Kaspi}, {Landecker}, {Leung}, {Li}, {Lin}, {Masui}, {McKinven}, {Mena-Parra}, {Merryfield}, {Meyers}, {Michilli}, {Milutinovic}, {Mirhosseini}, {M{\"u}nchmeyer}, {Naidu}, {Newburgh}, {Ng}, {Patel}, {Pen}, {Pinsonneault-Marotte}, {Pleunis}, {Quine}, {Rafiei-Ravandi}, {Rahman}, {Ransom}, {Renard}, {Sanghavi}, {Scholz}, {Shaw}, {Shin}, {Siegel}, {Singh}, {Smegal}, {Smith}, {Stairs}, {Tan}, {Tendulkar}, {Tretyakov}, {Vanderlinde}, {Wang}, {Wulf}, \& {Zwaniga}}]{CHIME2020}
{CHIME/FRB Collaboration}, {Andersen}, B.~C., {Bandura}, K.~M., {et~al.} 2020, \nat, 587, 54, \dodoi{10.1038/s41586-020-2863-y}

\bibitem[{{Cordes} \& {Lazio}(2002)}]{Cordes2002}
{Cordes}, J.~M., \& {Lazio}, T.~J.~W. 2002, arXiv e-prints, astro, \dodoi{10.48550/arXiv.astro-ph/0207156}

\bibitem[{{Daugherty} \& {Harding}(1996)}]{Daugherty1996}
{Daugherty}, J.~K., \& {Harding}, A.~K. 1996, \apj, 458, 278, \dodoi{10.1086/176811}

\bibitem[{{Duncan} \& {Thompson}(1992)}]{Duncan1992}
{Duncan}, R.~C., \& {Thompson}, C. 1992, \apjl, 392, L9, \dodoi{10.1086/186413}

\bibitem[{{Gaensler}(2014)}]{Gaensler2014}
{Gaensler}, B.~M. 2014, GRB Coordinates Network, 16533, 1

\bibitem[{{Gavriil} {et~al.}(2008){Gavriil}, {Gonzalez}, {Gotthelf}, {Kaspi}, {Livingstone}, \& {Woods}}]{Gavriil2008}
{Gavriil}, F.~P., {Gonzalez}, M.~E., {Gotthelf}, E.~V., {et~al.} 2008, Science, 319, 1802, \dodoi{10.1126/science.1153465}

\bibitem[{{Good} \& {Chime/Frb Collaboration}(2020)}]{Good2020}
{Good}, D., \& {Chime/Frb Collaboration}. 2020, The Astronomer's Telegram, 14074, 1

\bibitem[{{Han} {et~al.}(2021){Han}, {Wang}, {Wang}, {Wang}, {Zhou}, {Sun}, {Yan}, {Su}, {Jing}, {Chen}, {Gao}, {Hou}, {Xu}, {Lee}, {Wang}, {Jiang}, {Xu}, {Yan}, {Gan}, {Guan}, {Huang}, {Jiang}, {Li}, {Men}, {Sun}, {Wang}, {Wang}, {Wang}, {Xie}, {Xu}, {Yao}, {You}, {Yu}, {Yuan}, {Yuen}, {Zhang}, \& {Zhu}}]{Han2021}
{Han}, J.~L., {Wang}, C., {Wang}, P.~F., {et~al.} 2021, Research in Astronomy and Astrophysics, 21, 107, \dodoi{10.1088/1674-4527/21/5/107}

\bibitem[{Han {et~al.}(2025)Han, Zhou, Wang, Su, Yan, Jing, Yang, Wang, Wang, Xu, Cai, Sun, Yang, Xu, Wang, \& You}]{Han2025}
Han, J.~L., Zhou, D.~J., Wang, C., {et~al.} 2025, Research in Astronomy and Astrophysics, 25, 014001, \dodoi{10.1088/1674-4527/ada3b7}

\bibitem[{{Hibschman} \& {Arons}(2001)}]{Hibschman2001}
{Hibschman}, J.~A., \& {Arons}, J. 2001, \apj, 560, 871, \dodoi{10.1086/323069}

\bibitem[{{Hotan} {et~al.}(2004){Hotan}, {van Straten}, \& {Manchester}}]{Hotan2004}
{Hotan}, A.~W., {van Straten}, W., \& {Manchester}, R.~N. 2004, \pasa, 21, 302, \dodoi{10.1071/AS04022}

\bibitem[{{Hu} {et~al.}(2024){Hu}, {Narita}, {Enoto}, {Younes}, {Wadiasingh}, {Baring}, {Ho}, {Guillot}, {Ray}, {G{\"u}ver}, {Rajwade}, {Arzoumanian}, {Kouveliotou}, {Harding}, \& {Gendreau}}]{Hu2024}
{Hu}, C.-P., {Narita}, T., {Enoto}, T., {et~al.} 2024, \nat, 626, 500, \dodoi{10.1038/s41586-023-07012-5}

\bibitem[{{Israel} {et~al.}(2016){Israel}, {Esposito}, {Rea}, {Coti Zelati}, {Tiengo}, {Campana}, {Mereghetti}, {Rodriguez Castillo}, {G{\"o}tz}, {Burgay}, {Possenti}, {Zane}, {Turolla}, {Perna}, {Cannizzaro}, \& {Pons}}]{Israel2016}
{Israel}, G.~L., {Esposito}, P., {Rea}, N., {et~al.} 2016, \mnras, 457, 3448, \dodoi{10.1093/mnras/stw008}

\bibitem[{{Jiang} {et~al.}(2020){Jiang}, {Tang}, {Hou}, {Liu}, {Kr{\v{c}}o}, {Qian}, {Sun}, {Ching}, {Liu}, {Duan}, {Yue}, {Gan}, {Yao}, {Li}, {Pan}, {Yu}, {Liu}, {Li}, {Peng}, {Yan}, \& {FAST Collaboration}}]{Jiang2020}
{Jiang}, P., {Tang}, N.-Y., {Hou}, L.-G., {et~al.} 2020, Research in Astronomy and Astrophysics, 20, 064, \dodoi{10.1088/1674-4527/20/5/64}

\bibitem[{{Kaspi} \& {Beloborodov}(2017)}]{Kaspi2017}
{Kaspi}, V.~M., \& {Beloborodov}, A.~M. 2017, \araa, 55, 261, \dodoi{10.1146/annurev-astro-081915-023329}

\bibitem[{{Kirsten} {et~al.}(2021){Kirsten}, {Snelders}, {Jenkins}, {Nimmo}, {van den Eijnden}, {Hessels}, {Gawro{\'n}ski}, \& {Yang}}]{Kirsten2021}
{Kirsten}, F., {Snelders}, M.~P., {Jenkins}, M., {et~al.} 2021, Nature Astronomy, 5, 414, \dodoi{10.1038/s41550-020-01246-3}

\bibitem[{{Kramer} {et~al.}(2007){Kramer}, {Stappers}, {Jessner}, {Lyne}, \& {Jordan}}]{Kramer2007}
{Kramer}, M., {Stappers}, B.~W., {Jessner}, A., {Lyne}, A.~G., \& {Jordan}, C.~A. 2007, \mnras, 377, 107, \dodoi{10.1111/j.1365-2966.2007.11622.x}

\bibitem[{{Krimm} {et~al.}(2020){Krimm}, {Lien}, {Page}, {Palmer}, {Tohuvavohu}, \& {Neil Gehrels Swift Observatory Team}}]{Krimm2020}
{Krimm}, H.~A., {Lien}, A.~Y., {Page}, K.~L., {et~al.} 2020, GRB Coordinates Network, 28187, 1

\bibitem[{{Leahy} \& {Tian}(2008)}]{Leahy2008}
{Leahy}, D.~A., \& {Tian}, W.~W. 2008, \aap, 480, L25, \dodoi{10.1051/0004-6361:20079149}

\bibitem[{{Levin} {et~al.}(2010){Levin}, {Bailes}, {Bates}, {Bhat}, {Burgay}, {Burke-Spolaor}, {D'Amico}, {Johnston}, {Keith}, {Kramer}, {Milia}, {Possenti}, {Rea}, {Stappers}, \& {van Straten}}]{Levin2010}
{Levin}, L., {Bailes}, M., {Bates}, S., {et~al.} 2010, \apjl, 721, L33, \dodoi{10.1088/2041-8205/721/1/L33}

\bibitem[{{Li} {et~al.}(2021){Li}, {Lin}, {Xiong}, {Ge}, {Li}, {Li}, {Lu}, {Zhang}, {Tuo}, {Nang}, {Zhang}, {Xiao}, {Chen}, {Song}, {Xu}, {Liu}, {Jia}, {Cao}, {Qu}, {Zhang}, {Gu}, {Liao}, {Zhao}, {Tan}, {Nie}, {Zhao}, {Zheng}, {Zheng}, {Luo}, {Cai}, {Li}, {Xue}, {Bu}, {Chang}, {Chen}, {Chen}, {Chen}, {Chen}, {Chen}, {Cui}, {Cui}, {Deng}, {Dong}, {Du}, {Fu}, {Gao}, {Gao}, {Gao}, {Gu}, {Guan}, {Guo}, {Han}, {Huang}, {Huo}, {Jiang}, {Jiang}, {Jin}, {Jin}, {Kong}, {Li}, {Li}, {Li}, {Li}, {Li}, {Li}, {Li}, {Liang}, {Liu}, {Liu}, {Liu}, {Liu}, {Liu}, {Lu}, {Lu}, {Luo}, {Ma}, {Meng}, {Ou}, {Sai}, {Shang}, {Song}, {Sun}, {Tao}, {Wang}, {Wang}, {Wang}, {Wang}, {Wang}, {Wen}, {Wu}, {Wu}, {Wu}, {Xiao}, {Xu}, {Yang}, {Yang}, {Yang}, {Yang}, {Yi}, {Yin}, {You}, {Zhang}, {Zhang}, {Zhang}, {Zhang}, {Zhang}, {Zhang}, {Zhang}, {Zhang}, {Zhang}, {Zhang}, {Zhang}, {Zhang}, {Zhang}, {Zhang}, {Zhang}, {Zhang}, {Zhou}, {Zhou}, {Zhu}, {Zhu}, \& {Zhuang}}]{Li2021}
{Li}, C.~K., {Lin}, L., {Xiong}, S.~L., {et~al.} 2021, Nature Astronomy, 5, 378, \dodoi{10.1038/s41550-021-01302-6}

\bibitem[{{Lin} {et~al.}(2020){Lin}, {Zhang}, {Wang}, {Gao}, {Guan}, {Han}, {Jiang}, {Jiang}, {Lee}, {Li}, {Men}, {Miao}, {Niu}, {Niu}, {Sun}, {Wang}, {Wang}, {Xu}, {Xu}, {Xu}, {Yang}, {Yang}, {Yu}, {Zhang}, {Zhang}, {Zhou}, {Zhu}, {Castro-Tirado}, {Dai}, {Ge}, {Hu}, {Li}, {Li}, {Li}, {Liang}, {Jia}, {Querel}, {Shao}, {Wang}, {Wang}, {Wu}, {Xiong}, {Xu}, {Yang}, {Zhang}, {Zhang}, {Zheng}, \& {Zou}}]{Lin2020}
{Lin}, L., {Zhang}, C.~F., {Wang}, P., {et~al.} 2020, \nat, 587, 63, \dodoi{10.1038/s41586-020-2839-y}

\bibitem[{{Lorimer} \& {Kramer}(2004)}]{Lorimer2004}
{Lorimer}, D.~R., \& {Kramer}, M. 2004, {Handbook of Pulsar Astronomy}, Vol.~4

\bibitem[{{Lower} {et~al.}(2020){Lower}, {Shannon}, {Johnston}, \& {Bailes}}]{Lower2020}
{Lower}, M.~E., {Shannon}, R.~M., {Johnston}, S., \& {Bailes}, M. 2020, \apjl, 896, L37, \dodoi{10.3847/2041-8213/ab9898}

\bibitem[{{Lu} {et~al.}(2024){Lu}, {Zhou}, {Wang}, {Shao}, {Li}, {Vink}, {Li}, \& {Chen}}]{Lu2024}
{Lu}, W.-J., {Zhou}, P., {Wang}, P., {et~al.} 2024, \apj, 963, 151, \dodoi{10.3847/1538-4357/ad27cf}

\bibitem[{{Majid} {et~al.}(2020){Majid}, {Pearlman}, {Prince}, {Enoto}, {Arzoumanian}, {Gendreau}, {Naudet}, {Kocz}, {Horiuchi}, {Harding}, {Ho}, \& {Kuiper}}]{Majid2020}
{Majid}, W.~A., {Pearlman}, A.~B., {Prince}, T.~A., {et~al.} 2020, The Astronomer's Telegram, 13988, 1

\bibitem[{{Mereghetti} {et~al.}(2015){Mereghetti}, {Pons}, \& {Melatos}}]{Mereghetti2015}
{Mereghetti}, S., {Pons}, J.~A., \& {Melatos}, A. 2015, \ssr, 191, 315, \dodoi{10.1007/s11214-015-0146-y}

\bibitem[{{Nan}(2006)}]{Nan2006}
{Nan}, R. 2006, Science in China: Physics, Mechanics and Astronomy, 49, 129, \dodoi{10.1007/s11433-006-0129-9}

\bibitem[{{Nan} {et~al.}(2011){Nan}, {Li}, {Jin}, {Wang}, {Zhu}, {Zhu}, {Zhang}, {Yue}, \& {Qian}}]{Nan2011}
{Nan}, R., {Li}, D., {Jin}, C., {et~al.} 2011, International Journal of Modern Physics D, 20, 989, \dodoi{10.1142/S0218271811019335}

\bibitem[{{Olausen} \& {Kaspi}(2014)}]{Olausen2014}
{Olausen}, S.~A., \& {Kaspi}, V.~M. 2014, \apjs, 212, 6, \dodoi{10.1088/0067-0049/212/1/6}

\bibitem[{{Ransom}(2011)}]{Ransom2011}
{Ransom}, S. 2011, {PRESTO: PulsaR Exploration and Search TOolkit}, Astrophysics Source Code Library, record ascl:1107.017

\bibitem[{{Rea} {et~al.}(2014){Rea}, {Vigan{\`o}}, {Israel}, {Pons}, \& {Torres}}]{Rea2014}
{Rea}, N., {Vigan{\`o}}, D., {Israel}, G.~L., {Pons}, J.~A., \& {Torres}, D.~F. 2014, \apjl, 781, L17, \dodoi{10.1088/2041-8205/781/1/L17}

\bibitem[{{Rea} {et~al.}(2010){Rea}, {Esposito}, {Turolla}, {Israel}, {Zane}, {Stella}, {Mereghetti}, {Tiengo}, {G{\"o}tz}, {G{\"o}{\u{g}}{\"u}{\c{s}}}, \& {Kouveliotou}}]{Rea2010}
{Rea}, N., {Esposito}, P., {Turolla}, R., {et~al.} 2010, Science, 330, 944, \dodoi{10.1126/science.1196088}

\bibitem[{{Ruderman} \& {Sutherland}(1975)}]{Ruderman1975}
{Ruderman}, M.~A., \& {Sutherland}, P.~G. 1975, \apj, 196, 51, \dodoi{10.1086/153393}

\bibitem[{{Sathyaprakash} {et~al.}(2024){Sathyaprakash}, {Rea}, {Coti Zelati}, {Borghese}, {Pilia}, {Trudu}, {Burgay}, {Turolla}, {Zane}, {Esposito}, {Mereghetti}, {Campana}, {G{\"o}tz}, {Ibrahim}, {Israel}, {Possenti}, \& {Tiengo}}]{Sathyaprakash2024}
{Sathyaprakash}, R., {Rea}, N., {Coti Zelati}, F., {et~al.} 2024, arXiv e-prints, arXiv:2401.08010, \dodoi{10.48550/arXiv.2401.08010}

\bibitem[{{Stamatikos} {et~al.}(2014){Stamatikos}, {Malesani}, {Page}, \& {Sakamoto}}]{Stamatikos2014}
{Stamatikos}, M., {Malesani}, D., {Page}, K.~L., \& {Sakamoto}, T. 2014, GRB Coordinates Network, 16520, 1

\bibitem[{{Sun} {et~al.}(2011){Sun}, {Reich}, {Reich}, {Xiao}, {Gao}, \& {Han}}]{Sun2011}
{Sun}, X.~H., {Reich}, P., {Reich}, W., {et~al.} 2011, \aap, 536, A83, \dodoi{10.1051/0004-6361/201117693}

\bibitem[{{Tang} {et~al.}(2021){Tang}, {Zhang}, {Dai}, {Li}, \& {Wu}}]{Tang2021}
{Tang}, Z., {Zhang}, S., {Dai}, S., {Li}, Y., \& {Wu}, X. 2021, arXiv e-prints, arXiv:2106.04821, \dodoi{10.48550/arXiv.2106.04821}

\bibitem[{{van Straten} \& {Bailes}(2011)}]{van2011}
{van Straten}, W., \& {Bailes}, M. 2011, \pasa, 28, 1, \dodoi{10.1071/AS10021}

\bibitem[{{Wang} {et~al.}(2019){Wang}, {Zhang}, {Chen}, \& {Xu}}]{Wang2019}
{Wang}, W., {Zhang}, B., {Chen}, X., \& {Xu}, R. 2019, \apj, 875, 84, \dodoi{10.3847/1538-4357/ab0e71}

\bibitem[{{Yao} {et~al.}(2017){Yao}, {Manchester}, \& {Wang}}]{Yao2017}
{Yao}, J.~M., {Manchester}, R.~N., \& {Wang}, N. 2017, \apj, 835, 29, \dodoi{10.3847/1538-4357/835/1/29}

\bibitem[{{Zhang} {et~al.}(2020){Zhang}, {Jiang}, {Men}, {Wang}, {Xu}, {Xu}, {Niu}, {Zhou}, {Guan}, {Han}, {Jiang}, {Lee}, {Li}, {Lin}, {Niu}, {Wang}, {Wang}, {Xu}, {Yu}, {Zhang}, \& {Zhu}}]{Zhang2020}
{Zhang}, C.~F., {Jiang}, J.~C., {Men}, Y.~P., {et~al.} 2020, The Astronomer's Telegram, 13699, 1

\bibitem[{{Zhong} {et~al.}(2020){Zhong}, {Dai}, {Zhang}, \& {Deng}}]{Zhong2020}
{Zhong}, S.-Q., {Dai}, Z.-G., {Zhang}, H.-M., \& {Deng}, C.-M. 2020, \apjl, 898, L5, \dodoi{10.3847/2041-8213/aba262}

\bibitem[{{Zhou} {et~al.}(2023){Zhou}, {Han}, {Xu}, {Wang}, {Wang}, {Wang}, {Jing}, {Chen}, {Yan}, {Su}, {Gan}, {Jiang}, {Sun}, {Wang}, {Wang}, {Wang}, {Xu}, \& {You}}]{Zhou2023}
{Zhou}, D.~J., {Han}, J.~L., {Xu}, J., {et~al.} 2023, Research in Astronomy and Astrophysics, 23, 104001, \dodoi{10.1088/1674-4527/accc76}

\bibitem[{{Zhou} {et~al.}(2014){Zhou}, {Chen}, {Li}, {Safi-Harb}, {Mendez}, {Terada}, {Sun}, \& {Ge}}]{zhou2014}
{Zhou}, P., {Chen}, Y., {Li}, X.-D., {et~al.} 2014, ApJL, 781, L16, \dodoi{10.1088/2041-8205/781/1/L16}

\bibitem[{{Zhou} {et~al.}(2020){Zhou}, {Zhou}, {Chen}, {Wang}, {Vink}, \& {Wang}}]{Zhou2020}
{Zhou}, P., {Zhou}, X., {Chen}, Y., {et~al.} 2020, \apj, 905, 99, \dodoi{10.3847/1538-4357/abc34a}

\bibitem[{{Zhu} {et~al.}(2023){Zhu}, {Xu}, {Zhou}, {Lin}, {Wang}, {Wang}, {Zhang}, {Niu}, {Chen}, {Li}, {Meng}, {Lee}, {Zhang}, {Feng}, {Ge}, {G{\"o}{\u{g}}{\"u}{\c{s}}}, {Guan}, {Han}, {Jiang}, {Jiang}, {Kouveliotou}, {Li}, {Miao}, {Miao}, {Men}, {Niu}, {Wang}, {Wang}, {Xu}, {Xu}, {Xue}, {Yang}, {Yu}, {Yuan}, {Yue}, {Zhang}, \& {Zhang}}]{Zhu2023}
{Zhu}, W., {Xu}, H., {Zhou}, D., {et~al.} 2023, Science Advances, 9, eadf6198, \dodoi{10.1126/sciadv.adf6198}

\end{thebibliography}

\end{document}